\documentclass[acmsmall]{acmart} 
\AtBeginDocument{%
  }
\setcopyright{cc}
\setcctype{by}
\acmJournal{PACMHCI}
\acmYear{2025} \acmVolume{9} \acmNumber{7} \acmArticle{CSCW362}
\acmMonth{11} \acmPrice{}\acmDOI{10.1145/3757543}
\begin{document}
\title[A Node on the Constellation]{A Node on the Constellation: The Role of Feminist Makerspaces in Building and Sustaining Alternative Cultures of Technology Production}


\author{Erin Gatz}
\email{egatz@andrew.cmu.edu}
\affiliation{%
  \institution{Prototype PGH and Carnegie Mellon University}
  \city{Pittsburgh}
  \state{PA}
  \country{USA}
}

\author{Yasmine Kotturi}
\email{kotturi@umbc.edu}
\affiliation{%
  \institution{University of Maryland Baltimore County}
  \city{Baltimore}
  \state{PA}
  \country{USA}
}

\author{Andrea Afua Kwamya}
\email{andrea@bgomconsulting.com}
\affiliation{%
  \institution{Prototype PGH}
  \city{Los Angeles}
  \state{CA}
  \country{USA}
}

\author{Sarah Fox}
\email{sarahf@andrew.cmu.edu}
\affiliation{%
  \institution{Carnegie Mellon University}
  \city{Pittsburgh}
  \state{PA}
  \country{USA}
}

\renewcommand{\shortauthors}{Erin Gatz, Yasmine Kotturi, Andrea Afua Kwamya, and Sarah Fox}
\newcommand {\yasmine}[1]{{\textcolor{magenta}{#1}}}
\newcommand {\change}[1]{{\textcolor{black}{#1}}}

\begin{abstract}\change{
Feminist makerspaces offer community-led alternatives to dominant tech cultures by centering care, mutual aid, and collective knowledge production. While prior CSCW research has explored their inclusive practices, less is known about how these spaces sustain themselves over time. Drawing on interviews with 18 founders and members across 8 U.S. feminist makerspaces \textcolor{black} {as well as} autoethnographic reflection, we examine the organizational and relational practices that support long-term endurance. We find that sustainability is not achieved through growth or institutionalization, but through care-driven stewardship, solidarity with local justice movements, and shared governance. These social practices position feminist makerspaces as prefigurative counterspaces—sites that enact, rather than defer, feminist values in everyday practice. This paper offers empirical insight into how feminist makerspaces persist amid structural precarity, and highlights the forms of labor and coalition-building that underpin alternative sociotechnical infrastructures.}\end{abstract}




\received{October 2024}
\received[revised]{April 2025}
\received[accepted]{August 2025}

\maketitle

\section{Introduction}
\begin{quote} 
    “[Our makerspace] can only ever be a node on the constellation of projects happening in [our city]. How it fits within social justice and racial justice projects is about relationships between people at [makerspace] who have moved on. It’s an ecosystem sensibility.”  

    \hspace*{2em}-Shreya\footnote{All names used throughout this paper are pseudonyms, in the interest of preserving interviewees' privacy.}, Feminist Makerspace Co-Founder
\end{quote}

\change{Over the past two decades, feminist makerspaces have emerged across the United States as vital alternatives to the dominant logics of technology production \cite{foster2019claims, fox2015hacking, toupin2014feminist}. These spaces were founded in response to the exclusionary and extractive cultures of the tech industry, offering community-led infrastructures for technical practice, political organizing, and care work \cite{savic2018feminist}. Often self-organized and volunteer-run, feminist makerspaces seek not only to democratize access to tools and training, but to build collective spaces where people marginalized by racism, sexism, and classism can imagine and enact different technological futures.}

\change{Existing CSCW and HCI research has richly documented the inclusive pedagogical practices and design imaginaries of these spaces ~\cite{fox2015hacking, hedditch2021gendered, capel2020wooden, capel2021making, hedditch2023design}. These studies have shown how feminist values such as care, accessibility, and community accountability are woven into the everyday practices of making and learning in these spaces. Yet despite their transformative promise, many feminist makerspaces face significant challenges: managing volunteer burnout, maintaining leadership transitions, organizational instability, and the structural pressures of professionalization and growth. The closure of three prominent feminist makerspaces between 2014 and 2018 underscores these tensions and raises pressing questions about what enables some to endure while others shutter after only a few years.}

\change{With this research we ask, "What relational, organizational, and infrastructural practices enable feminist makerspaces to persist as counterinstitutions, and how do they model alternative cultures of technological work?" To explore this question, we draw on  semi-structured interviews with 8 founders and 10 members from \textcolor{black} {8} feminist makerspaces, cumulatively representing nearly 100 years of operation. These interviews were complemented by autoethnographic reflections from the first author, who co-founded and continues to help steward a feminist makerspace in the Rustbelt region of the United States.}

\change{Conversations with founders and members show that while funders often pressure feminist makerspaces to scale, those that persist prioritize deepening local relationships over organizational growth. This focus on community connection emerges as key to their long-term sustainability. Still, financial precarity and emotional burnout—often stemming from interpersonal tensions—remain major reasons for closure. To address these challenges, many turn to cooperative models that support multiple revenue streams and ongoing ties with former members. Sliding-scale fees, resident artist rent, grants, and long-term compensated staff help maintain institutional stability, enabling a wide range of vibrant programs. These practices position feminist makerspaces within a broader movement for cooperative work and alternative knowledge production. We argue that a solidarity-driven approach~\cite{dillahunt2023eliciting} is central not only to sustaining these spaces but also to supporting the collective survival of women \textcolor{black} {and people with marginalized gender identities} in tech. This ethos often takes shape through alignment with social and political causes, collaboration with justice-oriented groups, and an explicit commitment to marginalized communities—building a foundation for long-term resilience.}

\change{This paper makes three contributions to the CSCW community. First, we examine how feminist makerspaces practice long-term stewardship through distributed leadership, rotating governance, and affective labor. Second, we show how they resist the dominant logic of institutional growth, choosing instead to center solidarity with local communities as a foundation for sustainability. Third, we position these spaces as prefigurative counterinstitutions: not simply resisting exclusionary tech cultures, but actively building new ones grounded in feminist values of care, reciprocity, and collective imagination \cite{bardzell2018utopias}. This inquiry is especially urgent given the ongoing marginalization of women and people of color in the tech sector~\cite{weforumGenderRaceTech,OMara2022}, and the increasing precarity of equity work across institutional settings. As companies retreat from DEI commitments and academic programs face growing political scrutiny \cite{trump2025ending, techcrunchHereTech, wuschitz2022feminist}, community-led initiatives such as feminist makerspaces remain essential yet under-resourced infrastructures of justice. Understanding how these spaces persist can illuminate broader strategies for building and sustaining counterinstitutions amid such "hostile ecologies" \cite{tandon2022hostile}.}

\section{Related Work}
This paper is situated between two key bodies of scholarship that motivate and animate the study. First, we synthesize the history and role of feminist makerspaces as sites of alternative tech production, especially by centering marginalized voices within tech. Then, we discuss CSCW scholarship on the role of solidarity economies, prefigurative design, and counterspaces to understand how feminist makerspaces not only resist dominant tech cultures, but actively build and sustain community-rooted alternatives. 

\subsection{Feminist HCI, Makerspaces and Alternative Cultures of Tech Production}
Community-based spaces such as hackerspaces, makerspaces, and open source technology conferences produce software, hardware, and technical knowledge that seek to disrupt the status quo~\cite{wuschitz2014hacking, Dunbar-Hester_2020}. Operating outside of institutional settings such as tech companies and universities, these spaces create alternative pathways into technological practice. \change{For example, D'Ignazio et al. \cite{d2023data} describe their transformation of a hackathon into a feminist tech festival 'Make the Breast Pump Not Suck', which included a zine library, a baby-friendly space, a policy summit, and an art exhibition. Similarly, during the COVID-19 pandemic, Hedditch and Vyas \cite{hedditch2021gendered} documented an online hub where immigrants and refugee members of a feminist makerspace provided each other with social support and participated in remote making tutorials. These examples illustrate how such spaces reconfigure participation in technology by foregrounding care and accessibility.}

Feminist makerspaces explicitly identify with feminist principles, offering supportive environments for women and people with marginalized gender identities to work on technical projects~\cite{toupin2014feminist}. Unlike many mainstream makerspaces, they aim not just to democratize access to tools, but to challenge the norms and exclusions embedded in tech culture~\cite{roldan2018university}. As Costanza-Chock argues, most makerspaces fail to disrupt systemic oppression because they assume that “providing people with tools is enough”~\cite{costanza2020design}. In contrast, feminist makerspaces invest in learning pathways that honor the experiences and knowledge of marginalized communities. Capel et al.~\cite{capel2020wooden} provide an illustrative case in Australia where participants engaged in woodworking while sharing personal stories of exclusion from making practices. Themes such as housing insecurity, health challenges, and relational dynamics emerged, revealing how feminist makerspaces create spaces where storytelling and craft can intersect. In doing so, participants can make historically masculine practices more accessible. 

Feminist tech spaces also challenge dominant narratives of techno-heroism, individualistic, novelty-driven innovation, by emphasizing relational, community-sustaining practices~\cite{balsamo2020designing}. Fox et al.~\cite{fox2015hacking} frame feminism as a lens to interrogate the widespread pressure to adopt new technologies for their own sake. They document how the activities supported by feminist makerspaces  resist this logic and revalue nondigital craft work. As Rogers~\cite{rogersmaking} notes, these spaces reveal a long history of marginalized communities as everyday innovators and survivalist hackers. Making in this context can include knitting circles as a radical craft, protest banner design, or circuit boards soldered from recycled materials. Here, innovation is redefined as sustaining life, not scaling products.

Toombs et al.~\cite{toombs2015proper} extend this view by examining the often invisible maintenance work that sustains hackerspaces. They highlight how acts of care, both explicit and implicit, are essential to the longevity of these communities. Yet, such relational labor is rarely centered in research on technology production. \change{We build on this work by offering a detailed account of how maintenance and care are fundamental to the resilience of feminist makerspaces. Importantly, focusing on the needs of marginalized communities is not only a matter of justice; it is a strategy for organizational resilience. Feminist makerspaces distribute responsibility across diverse constituencies, cultivating cooperative structures that do not depend on singular forms of support. The present study extends existing scholarship by providing empirical and autoethnographic insights into how feminist makerspaces support the cohesion of social constellations and how they sustain alternative cultures of technology production through care, community, and collective imagination.}

\subsection{Solidarity Economies, Prefigurative Design, and Counterspaces}
CSCW and HCI scholars have increasingly turned to frameworks such as prefigurative design~\cite{asad2019prefigurative}, counterspaces \cite{to2023flourishing, silva2024or, brooks2021uplifting}, and solidarity economies~\cite{vlachokyriakos2017hci, spade2020mutual, dillahunt2023eliciting} to understand how communities build alternative sociotechnical infrastructures that resist dominant institutional logics. \change{Prefigurative design, in particular, calls for design practices that embody the values of the just worlds they seek to bring about, rather than deferring equity until after implementation. These approaches foreground relational practices such as mutual aid, distributed governance, and situated accountability, offering a lens for understanding how grassroots organizations act on justice in everyday settings.}

\change{Within CSCW, counterspaces have emerged as a powerful concept for understanding how marginalized groups create environments of affirmation, resistance, and alternative knowledge production. Rankin et al. \cite{rankin2025}, for instance, describe how Black women in computer science education cultivate counterspaces that enable them to process structural harm and generate collective survival strategies. Erete et al. \cite{erete2021} similarly examine how Black girls co-design civic technologies within trusted, affirming environments, allowing them to explore futures otherwise foreclosed in dominant STEM spaces. These counterspaces are not neutral backdrops, but actively shape participants’ political identities and design imaginaries \cite{Tunstall_Agi_2023}. Other CSCW work has examined how community tech initiatives operate as counter-institutional infrastructures. Dillahunt et al. \cite{dillahunt2023eliciting} explore Afrofuturist design practices among working-class Detroiters, revealing how speculative design workshops became spaces for articulating collective economic alternatives rooted in shared struggle. Similarly, Vlachokyriakos et al. \cite{vlachokyriakos2018infrastructuring} trace how solidarity networks in Greece mobilized digital tools to organize mutual aid, food distribution, and refugee housing—offering both sociotechnical infrastructure and community governance models distinct from state or market institutions.}

\change{CSCW scholars have also examined cooperative responses to extractive platform economies. Tandon et al. \cite{tandon2022hostile} describe how creators of worker-led alternatives to gig platforms must navigate “hostile ecologies,” where sustaining community trust, technical functionality, and legal legitimacy requires ongoing infrastructural and emotional labor. Calacci and Monroy-Hernández \cite{calacci2025fairfare} study participatory platforms that help rideshare drivers surface algorithmic harm, highlighting how such tools gain power when embedded in broader organizing efforts. These projects illustrate how counterspaces and countertechnologies are deeply relational—dependent on social trust, shared commitments, and sustained coalition work.}


\change{This paper builds on this literature by examining how feminist makerspaces operate as long-term counterspaces embedded within regional solidarity ecosystems. Although prior work has theorized the importance of counter-institutional design and collective care, fewer studies have focused on the organizational practices that sustain such efforts over time. By focusing on feminist makerspaces as sites of ongoing negotiation between prefigurative values and material constraints, this work contributes to CSCW conversations on how alternative tech cultures persist in the face of burnout, institutional cooptation, and resource scarcity. In particular, this paper considers how feminist principles, grounded in intersectionality, mutual aid, and local accountability, are operationalized not only in moments of resistance, but in the maintenance and endurance of community-led infrastructures. In doing so, this work deepens understandings of how counterspaces function not simply as reactive shelters, but as proactive and evolving sites of feminist worldbuilding.}


\section{Methods} We interviewed founders and members of feminist makerspaces throughout the United States using \change{a snowball sampling method to recruit participants}. To contextualize and expand our interpretations, we also drew from the first author’s experience of founding a feminist makerspace, using this experience as an object of study through autoethnographic analysis. By engaging both the insights of others and our personal stories, we assert that lived experiences, our own and those of our participants, are valid sources of knowledge and expertise \cite{rapp2018autoethnography}.

\subsection{Participants and Recruitment}
We began reaching out to feminist makerspace founders and members in March 2020 just as the COVID-19 pandemic unfolded. We focused our outreach on organizations that describe themselves as makerspaces or hackerspaces and that have feminist values in their mission statements. Some of the participants we interviewed described their spaces as self-organized collectives that support women artists. Other makerspace participants stated that they created their space in response to the questions ``Where are the women, the people of color, the queer folk, the blue collar workers? Where are our trusted allies?'' However, other spaces have the explicit goal of fostering a creative safe space for women and nonbinary people, while being ``people-of-color-led, gender-diverse, queer, and trans inclusive''. Although there are differences in how these makerspace participants self-identify and describe their activities, they all share the common goal of providing shared space, tools, and skills for community members in a feminist environment.

We recruited participants by emailing makerspace founders (some email addresses were listed on the makerspace websites, and other email addresses were given to us by people we know who have been members in those makerspaces). For example, some of the makerspace members at a local organization had previously lived in other cities and had been members at other makerspaces, so they had personal connections with those makerspace founders and members. Even so, \textcolor{black} {some} makerspace founders were reluctant to meet to discuss the sustainability and/or closure of their makerspace due to the soreness of the subject (especially if their space had closed). However, the first author, as a cofounder of a feminist makerspace herself, was able to connect with founders over their shared struggles and lived experience.

\change{In total, we reached out to 37 people and 18 people agreed to participate in an interview.} The interview participants included 8 founders and 10 current members of feminist makerspaces and hackerspaces from three regions of the United States: the Bay Area, Pacific Northwest, and North East (see Table~\ref{tab:participant-table}). Organizations were founded between 2000 and 2016, with three of the organizations having closed and five of the organizations still open during the writing of this paper in 2025. Thus, during the interviews, 15 participants reflected on their experience actively running an open makerspace, while 3 participants reflected on their experience with a makerspace that was no longer open at the time of the interview. Given that approximately one third of the feminist makerspaces we included in our study shut down after their first few years of operation, we wanted to explore the shared challenges that feminist makerspaces experience and how some spaces are able to overcome these challenges.

\begin{table}
    \centering
    \small
    \change{
        \begin{tabular}{|c|c|c|c|c|}
          \hline
        \textbf{Region} & \textbf{Makerspaces} & \textbf{Participant Roles} & \textbf{Tenure of Makerspaces}\\
          \hline
        North East & 2 makerspaces & 2 founders, 8 members & 2000-present; 2016-present\\
          \hline
        Pacific Northwest & 3 makerspaces & 3 founders & 2010-2014; 2013-2014; 2013-2018\\
          \hline
        Bay Area & 3 makerspaces & 3 founders, 2 members  & 2011-present; 2011-present; 2013-present\\
          \hline
    \end{tabular}
    \caption{Participant Roles and Feminist \textcolor{black} {Makerspace} Tenure as of 2025}
    \label{tab:participant-table}}
\end{table}

\subsection{Interviews and Data Analysis}
All interviews were conducted via a virtual meeting platform with automatic transcription software, lasting between 60 and 90 minutes. Each interview followed a protocol in which we introduced ourselves as a feminist makerspace cofounder and as researchers of feminist technologies. We told interviewees that we were conducting a study on the motivations, ideals, and activities of those who are organizing or participating in feminist makerspaces. We opened up each interview by asking the participants to describe their background, which we intentionally left open-ended. Participants discussed their academic background, careers, ethnicity and racial identities, gender identities, and economic background. We then asked what motivated them to get involved or start a makerspace with an explicitly feminist mission, and we asked participants to describe how their involvement has changed over time. We were also interested in stories of the challenges that feminist makerspace members have faced throughout their involvement, how they addressed these challenges, and we specifically asked participants ``If you could change one thing about the space, what would it be?'' We also asked interviewees to describe the values and ideals of their makerspaces and the degree to which they have seen racial justice enacted through the activities of the space. The interviews were not differentiated based on whether the participant's makerspace was still open at the time of the interview, and we asked the same questions of all the participants. 

To analyze our findings, we drew on interpretivist perspectives that aim to translate rather than transcribe social phenomena~\cite{huberman2014qualitative}. This process centered on collaborative qualitative analysis of the interview data, with the first author initially coding each transcript, followed by an iterative process in which the coauthors reviewed and refined the codes. Through weekly meetings over the course of a year, we collectively identified emergent themes, discussed patterns across interviews, and engaged in ongoing refinement of analytic memos~\cite{charmaz2006constructing}. \change{Our initial rounds of coding focused on what motivated people to start feminist makerspaces, the challenges faced by organizations to stay open, and how racial justice was reflected in the spaces' activities. Motivation to open feminist makerspaces stemmed largely from people experiencing hostile work environments in the tech sector, where they regularly faced sexism and racism (coded as 'racist work environment' or 'sexist work environment'). The main challenge faced by organizations was ensuring long-term sustainability without relying on corporations and foundations (coded as 'resisting institutional reliance' and 'community member driven'). Supporting marginalized community members to take on leadership roles in the organization was the most common example of racial justice activities in the spaces (coded as 'centering marginalized voices'). }

Importantly, we recontextualized these data through an autoethnographic lens, drawing on the first author's reflections on cofounding and managing a feminist makerspace for over eight years. \change{Autoethnography, as a method, enables researchers to connect personal experience with broader sociocultural and organizational dynamics, fostering a mode of inquiry that is both reflexive and situated \cite{ellis2011autoethnography}. As someone who established a nonprofit organization with an explicitly feminist mission and who has grappled with the ethical and strategic complexities of accepting corporate funding, the first author's experience provided a critical interpretive frame for analyzing our interview data. For example, upon learning that many feminist makerspace founders struggled with corporate and philanthropic entanglement, the first author wrote a reflection about how she and the cofounder of her feminist makerspace had initially accepted corporate funding and then later transitioned to a membership-driven model based on income-informed sliding scale fees. These autoethnographic reflections appear throughout our findings as a personal response to the themes that emerged from our interviews.}

\change{Rather than treating this personal history as anecdotal, we engaged it as a legitimate site of knowledge production—one that could be put into productive dialogue with participants’ narratives to reveal shared tensions, strategies, and contradictions in feminist organizing. This methodological orientation aligns with recent work in HCI and CSCW that centers autoethnography as a way to examine the gendered, affective, and infrastructural dimensions of making and community work \cite{hedditch2021gendered, rapp2018autoethnography}. By layering personal reflection with participant accounts, we sought to both validate the first author's lived experience and enrich our empirical analysis through reciprocal triangulation.}

\change{Through this iterative analysis drawing together interviews and autoethnographic reflections, we identified three interrelated practices that contribute to the endurance of feminist makerspaces: resisting reliance on institutional funding, refusing pressures to scale, and deepening relationships with neighboring justice organizations. Rather than viewing these practices as isolated strategies, we understand them as mutually reinforcing elements of a broader solidarity-based approach—one that privileges collective survival, community accountability, and relational forms of sustainability over institutional growth or technical innovation alone.}

\subsection{Positionality}

Our research is shaped by our varying relationships to feminist makerspaces as a central site of study, as well as by our institutional and disciplinary affiliations. \change{As Jackson et al. \cite{jackson2024positionality} argue, positionality statements should move beyond listing identities to meaningfully connect researchers’ social locations to their scholarly work. In this spirit, we articulate how our positions as researchers, practitioners, and community members informed our approach to this work.}

\change{Our team includes those with backgrounds in community organizing, technology design, media production, and feminist research, all of whom have participated in or studied feminist makerspaces in various capacities. Several of us have been directly involved in the organization of feminist makerspaces, allowing us to engage with these sites as both insiders and analysts. Notably, the first author previously worked at a franchised for-profit makerspace (founded by big tech actors and infused with federal funds) before cofounding a nonprofit feminist makerspace. Likewise, the second and third authors later became members and leaders of this space. These roles provided us with first-hand perspectives on the organizational, technical, and cultural dynamics that shape feminist tech communities. Our experiences navigating these spaces, whether through governance, education, or creative work, inform our analysis of the tensions between grassroots activism and institutional structures.}

\change{Collectively, we approach this research through a commitment to participatory inquiry and critical feminist analysis. Our perspectives are shaped by our engagements with racial and gender equity initiatives, as well as our positionalities as women working across academic and community settings. These experiences influence how we frame research questions, interpret findings, and navigate ethical responsibilities when studying feminist technology spaces.}

\section{Findings}
\change{In what follows, we organize our findings into three themes that emerged across interviews and autoethnographic reflection. These themes reflect how feminist makerspaces persist not by aligning with dominant models of nonprofit success, but by enacting alternative logics of governance, care, and coalition. First, we examine how feminist makerspaces navigate bureaucratic barriers and resource constraints without becoming dependent on corporate or philanthropic institutions. Second, we explore how leaders deliberately resist growth as a measure of success. Finally, we show how long-term sustainability is grounded in reciprocal relationships with neighboring communities and movements.}

\subsection{Bureaucratic Barriers and Fiscal Entanglements}
The inability of short-term investments to set organizations up for long-term financial success was a recurring theme throughout our interviews with founders. Among the founders we interviewed, many shared that initial donations from founding team members played an important role in securing leases, paying security deposits for spaces, and purchasing initial equipment. However, these initial investments often came at a long-term cost when founding members left the organization, leaving a financial gap to be filled. Such donations were typically made by members of the makerspace who worked in the tech sector, had sufficient disposable income, and did not expect to be paid back for their contribution. Members were motivated to underwrite the initiative because they wanted to create a space that felt safer and more supportive than their tech jobs, and they had the income to make it possible. For instance, \change{Elliott, a cofounder of a Bay Area-based space that is still open}, recalled:
\begin{quote}
“Three or four of us decided, if this goes south and no one joins, we can afford to pay this for a year.”    
\end{quote}

Here, Elliott emphasized the critical role that founding members often made in short-term investments to open up new feminist makerspaces. However, these investments were not intended to be long-term recurring contributions, and greatly depended on the resources founders had access to. \change{Gene, a cofounder of a feminist makerspace in the Pacific Northwest that has closed}, pointed to the broader structural pressures of gentrification and tech-driven development in their city:

\begin{quote}  ``As more money trickled into [the city], there were less spaces we could afford. It was hard to find a space that was accessible and easy to get to from public transit. I had a software job and was able to make a big investment.''
\end{quote}
 
The gentrification resulting from an influx of tech fueled a crisis of affordability for Gene's makerspace, and therefore they relied on their personal income to sustain the space. Despite intentions to avoid reliance on major personal or corporate contributions, most spaces in our study ultimately relied on them in their early stages or times of need. Short-term investments from founders were helpful in paying the first month's rent and deposit for a new space, purchasing makerspace equipment or furnishing the space. However, the long-term and recurring expenses associated with the legal entity required a broad community of committed donors that took time to establish. Such long-term, recurring expenses included equipment maintenance and repair, bookkeeping, cleaning services, materials and supplies, paying for annual tax filing, liability insurance, and directors/operators insurance. In addition to these material costs, there were also significant labor costs. Time-intensive tasks included coordinating a Board of Directors with all of its committees, meetings, and agendas. In short, starting a makerspace, like any other nonprofit organization, required upfront funding and labor. However, the long-term success of these nonprofits ultimately depended on the time investment, commitment, and resilience of a broad base of members.

This tension of needing to steward a legal entity in the long term while only having short-term resources was a frequent theme in our interviews with founders and members. \change{Marina, a cofounder of a North East-based space that is still open}, spoke about the turnover inherent among the Board of Directors:

\begin{quote}
    ``It used to be four formal roles on the steering committee: membership, finance, programming, operations. But it was really hard because it ended up being a lot of turnover...once something becomes more established, it’s like people don’t want to be part of it anymore. I think a lot of how [this makerspace] functions is like an incubator, where people realize what they want to be a part of, and then they go somewhere else to focus on it.''
\end{quote}
 
What Marina described is the initial excitement and energy that surrounds starting a new project or founding a new space, which can then dwindle when founding members are faced with the often bureaucratic and tedious aspects of maintaining the organization over time. Furthermore, Marina spoke to the idea that her feminist makerspace does not serve as an end-all for members, but instead that it serves as an entry point and incubation period for people who care about equity, design, and tech. This entry point was often a low or no barrier for people to get involved in nonprofit organization management, and once members felt as though they were ready to move on, they joined other organizations to deepen their skillset. In other words, this feminist makerspace served as an incubator and on-ramp for participants to gain nonprofit management experience, as well as social justice organizing skills.

High turnover as a result of member burnout and founders’ excitement dwindling over time was another common theme among founders we spoke with whose spaces had closed. Related to this high turnover was the frequency at which many of the spaces recruited new members to join the Board of Directors; founding members recruited new members to join the board, even though many of these members were relatively unfamiliar with the organization and did not join with the intention of leading it. \change{Sam, a newly minted board member of a Bay Area space that is still open}, recalled:

\begin{quote}
``It was June last year when I joined and only a few months later, in September or October they were looking for new [board] members, a call went out on the listserv. As someone who works in nonprofit development, I work with board members and I’ve been curious about what it would be like to be on a board of an organization. Looking for a low financial commitment to get into that so it seemed like perfect timing. Did a couple meetings with the current board members, and they didn’t mind that I didn’t know what I was doing because that’s the spirit of this whole thing.''
\end{quote}
 
As Sam described, she had only been involved in the makerspace for three or four months when she was asked to join the Board of Directors. She agreed to do so because she saw this leadership opportunity as a way to learn more about what it means to serve on a Board of Directors for a nonprofit organization. Importantly, she also felt welcomed and encouraged to join despite not having the requisite knowledge. 

In addition, we found that among the spaces in this study that closed after a few years, there was a common theme of founders leaving shortly after opening the space, leaving the legal and financial responsibilities of running a nonprofit organization to new members who were unfamiliar with the role of the Board of Directors. Across our interviews, we observed that while the initial vision and momentum of these spaces was set by a team of founders, there was less cohesion in the organization if the founders quickly cycled out.

\subsection{Tensions of Growth and the Need to Resist Scaling}
Across our conversations, interviewees noted that focusing on the depth of relationships between founding members versus scaling the organization was critical to organizational sustainability. In particular, developing decision-making processes and slowing down to discuss shared values, as well as potential political misalignments among members, was key to the long-term success of these organizations. According to Elliott:

\begin{quote}``Our biggest problem is financial. It’s hard to raise money when we don’t want to charge people to use the space. Our biggest challenges are on finances and capacity for people to do the organizing. We don’t want to ask Google or Facebook for money.''
\end{quote}
 
This tension-between wanting to provide a community resource at an affordable cost while also not wanting to accept corporate donations or philanthropic funding-was a common theme in our interviews with founders. In addition to not wanting to accept corporate funding, there were capacity questions as well. Who would write the grant applications? Who would meet with funders to provide quarterly and mid-term updates? Who would go to the foundation gatherings and networking events? Instead of having a formal fundraising person on staff, the participants we spoke with emphasized the importance of the space being financially accessible to members and using a sliding scale fee model for community members to use the space, particularly for progressive political meetings and community events.

\change{\textbf{\textit{Autoethnographic reflection.}} During the first author's experience in founding and maintaining a feminist makerspace for eight years, she had to wear many hats to conduct the necessary stewardship required to maintain the space. This unpaid labor quickly led to burnout. For example, during the first five years of running a feminist makerspace in the North East that is still open, the first author was responsible for a wide range of operational tasks such as paying bills, updating the website, coordinating open houses, maintaining equipment, taking out the trash, sweeping floors, and other similar tasks that required daily and weekly upkeep. Not only do these tasks require daily attention to detail (e.g., "Is there toilet paper in the bathroom?" "Is the laser cutter bed vacuumed out?" "Is the 3D printer bed calibrated?" "Has the rent been paid?"), but they can also bring up organizational issues ("What do we do if someone sleeps in the space overnight?" "What do we do if our pro-choice posters cause an argument?" "Do we raise money for an elevator so that our building can be wheelchair accessible even though we do not own the building, because the landlord will not do it?" "Who took the sewing machine home and did not return it?") Each of these issues are persistent and complex tensions that exist in feminist makerspaces. In fact, in the first author's own experience, this unpaid stewarding work led to the occasional need to take a personal leave of absence from the organization.}

\subsubsection{Pressure to Always ``Grow, Grow, Grow''}
\change{One potential avenue for organizational sustainability and financial stability identified by many makerspace participants was through growing the membership base by offering tiered pricing based on income, such that a larger number of people could afford access to the space. In fact, all of the participants we interviewed wanted to establish a broad constituent base that could sustain the space through a collective fundraising approach.} Additionally, small grants from foundations or government might be part of this collective approach, but many participants wanted grants to be supplementary revenue rather than having the space reliant on these sources.

Among the founders we spoke with, there were many ideas around how feminist makerspaces could be sustainable through such a collective approach. \change{Shreya, a cofounder of a long-running feminist makerspace in the North East that is still open}, discussed the importance of community-level relationships in ensuring the financial stability of the space:

\begin{quote}``We don’t want to get any bigger and we don’t want to do any more. If it’s bigger, it should be about depth of relationships...There’s a lot of pressure in the nonprofit community to always grow, grow, grow. But you can say, no that’s enough. Whatever comes next has to be about deepening and strengthening.''
\end{quote}

In this particular North East-based makerspace, financial sustainability was achieved when the makerspace decided to stop expanding its outreach and programs, and instead focus on paying its two cofounders and investing in its artists-in-residence. This was primarily achieved by renting out its building to artists as subtenants. Engaging artists as subtenants was also used as a strategy to resist gentrification in the neighborhood because subtenants often paid below market rates for rent.

Relatedly, in \change{a Bay Area makerspace that is still open}, founding members wanted a creative, shared workspace and low-income residents in the surrounding neighborhood wanted affordable access to its programs and equipment. According to \change{Tye, a cofounder of this space}: 

\begin{quote}
``We moved out of our space [in a central neighborhood] and are now on the outskirts in a neighborhood with complicated racial dynamics. There is public housing where parts of it are being torn down and being rebuilt. It’s a somewhat industrial neighborhood but it’s [an expensive city] so the nonpublic housing is wildly expensive.''
\end{quote}

\change{Tye described their makerspace as being located in a low-income neighborhood where residents are at risk of displacement. At the same time, Tye expressed concern that early investments, whether from tech companies or personal contributions, could potentially contribute to gentrification of the neighborhood surrounding the makerspace. Although the neighborhoods in which many makerspaces opened were initially affordable, we observed a theme of makerspace participants voicing concerns about how their presence could impact the surrounding communities over time, particularly by raising rents for their neighbors. In spaces that persisted over time, these concerns often led to a tiered membership fee structure based on income, and outreach efforts that centered on engaging marginalized community members, and in particular low-income people of color.}

Relatedly, Shreya discussed the importance of centering people of color in the space and taking direction from communities of color in the neighborhood, particularly as a way to center their needs and to ensure that they felt comfortable and represented in the makerspace:

\begin{quote}``What cultural shifts need to happen in order to switch over from being a space where there are some POC who feel comfortable and have made projects here to investing in projects in our community that are POC led?''
\end{quote}
 
Rather than simply being a space that was inclusive for people of color or that had racially diverse membership, Shreya emphasized the importance of stepping outside of the space to take direction from POC-led projects in the neighborhood and to support their work on racial justice projects, which ranged from affordable housing campaigns to workforce development programs. This was important not only for sustaining the longevity of the makerspace but also for ensuring that its resources were accessible and relevant to communities of color. Having relationships with their low-income neighbors, and engaging low-income residents as makerspace members, allowed makerspaces to broaden their membership base while resisting institutional pressure to scale the organization. 

\change{\textbf{\textit{Autoethnographic reflection.}} In the first author’s experience of founding a feminist makerspace, a large grant was offered from a technology company in our first two years of operation. Because we needed the money and did not yet have collective decision-making processes in place, we decided to take the funding. Although this money allowed the space to be expanded, purchase equipment, and hire a staff person, it alienated some members who initially saw the makerspace as an anticapitalist and collective effort to reimagine tech. Women, people with marginalized gender identities and people of color had become involved in the space as a way to decompress and process their stressful and often sexist and racist workplaces. These workplaces were often large tech companies, financial institutions, and educational institutions which, at the same time as underpaying and underpromoting women and people of color, were funding community-based initiatives to support women and people of color. For some members, accepting funds from these institutions was seen as problematic as such efforts were viewed as publicity campaigns that would not serve to change discriminatory hiring and promotional processes. As our organization grew and developed, we stopped accepting corporate grants and instead focused on growing our membership base, and eventually it became our main source of revenue, providing enough to pay rent and part-time staff.}

\subsection{Deepening Relations with Neighbors and Reclaiming Making}
Across our interviews there was a central theme of makerspace members drawing on the expertise of local neighborhood residents, particularly for their lived experience with issues of equity and justice. For instance, \change{Elliott, a cofounder of a Bay Area space that is still open,} discussed how to do meaningful equity work in the space over the member email list:

\begin{quote}``There was a question on the list about what are we doing about diversity and inclusion. I wrote back saying, we are in this building which is very heavily POC led organizations. Why don’t we talk to our neighbors and invite them into our space?''
\end{quote}
 
In a shared building where there were multiple social justice organizations working alongside each other as tenants, collaborating with neighbors was often seen as the most direct way to create city-wide partnerships related to gender and racial justice. This Bay Area makerspace was located in a high-poverty area and next to a women’s shelter. As Elliott further recalled about one member of this space:

\begin{quote}
``Dana was homeless and living in a shelter, going to code school, and using [the feminist makerspace] as a springboard to do some outreach for the community. Dana is a Black trans woman. She was getting women to apply and come to the space, running events.''
\end{quote}
 
Because of Dana's lived experience as a low-income, Black, trans woman, who was also in code school, she was in a ideal position to recruit members with similar lived experience. Relatedly, Elliott realized that forging connections with their neighbors who were part of racial justice organizations led by POC would be a straightforward and sustainable approach to attracting members. However, not all valued Dana's approach of engaging with neighbors in one-on-one conversations. Elliott continued on to share another founding member's pushback to Dana's community engagement:

\begin{quote}
    ``[Another founding member] was like, `You’re not doing the role, I want to ask you to step down from the membership role.' [This founding member] wanted formality and a professional approach. It wasn’t in a space where she could see the work that Dana was doing. She didn’t see what talking to people one-on-one could do.'' 
\end{quote}

\textcolor{black} {Dana thought that forging connections with their neighbors would be a straightforward and sustainable approach to attracting members. But one of the founding members disagreed with this approach because she saw it as ‘unprofessional.’ Instead, this founding member wanted to hire outside consultants to give their advice about how the feminist makerspace could be more inclusive to marginalized communities.}

Conflicts surrounding member recruitment methods were addressed by some feminist makerspaces through formalized processes. In order to center the experiences of marginalized people in their spaces, many of the feminist makerspace participants we spoke with created codes of conduct about who could join the space. For example, rather than creating a space that is radically inclusive of anyone, one makerspace founder discussed the need to prioritize and explicitly welcome marginalized communities. \change{Tye, who is cofounder of both a feminist makerspace in the Pacific North West that is now closed, as well as cofounder of a feminist makerspace in the Bay Area that is still open, made the following point:}

\begin{quote}
``When I think about the [makerspace] code of conduct, it’s a shift in priorities from 'let’s be a cool hacker iconoclast and break stuff' to 'what is this space that we’re allowed to participate in, where people are underrepresented in other communities are explicitly welcomed?' That was the radical thing about feminist makerspaces and hackerspaces, was that fundamental shift. It’s not actually possible to include everyone, you can’t, so being deliberate about who you’re choosing to include.''
\end{quote}
 
Rather than creating a space that provided equal access to anyone, many of the makerspace participants we interviewed discussed the importance of creating a space with an equity approach to access. Centering the experiences of women, people with marginalized gender identities, and people of color is one way to decenter dominant, white, and heteronormative ways of producing technology. This emerged as a common theme across our interviews: many feminist makerspace participants explicitly welcomed discussion and interactive workshops around embodied ways of knowing. Instead of creating a race-neutral or gender-neutral space, feminist makerspace participants wanted space to explicitly address and explore the ways in which identities are intertwined with the world around us. This intersectional approach to understanding identity often challenged the gender binary and created a supportive space for transgender members. According to \change{Zoe, a cofounder of a Pacific North West space that is now closed:}

\begin{quote}
``Cis-feminism I’m not finding to be particularly useful for understanding trans issues. I transitioned at 30 and feel a different connection to different genders at different times. Like men are trash! But what about trans men?''
\end{quote}
 
Zoe spoke about their experience coming out as a trans man while managing a feminist makerspace and how their perspective on who should be centered in the work changed as well. Specifically, as they came out as a trans man, they no longer felt it was helpful to center the experiences of cis women in feminist makerspaces. Similarly, many of the participants we spoke with were grappling with how to challenge a gender binary that is inherent in larger societal structures and institutions while running a space that exists within this sociopolitical context. \change{In fact, this is where counterinstitutions come into practice: participants addressed this complexity by updating codes of conduct to center not just women but also people with marginalized gender identities. The codes of conduct included policies such as not assuming someone’s gender identity, acknowledging the privilege that one carries into a space, and establishing trans-led workshops and events for trans-identified participants.}

\subsubsection{Reclaiming Making and Hacking as a Way of Life}
We observed across our interviews that a collective survival approach rooted in solidarity with POC-led racial justice organizations can be one way to interrupt the white and macho tech world. According to one cofounder, \change{Elena, whose Bay Area space is currently open and led by people of color}:

\begin{quote}
``I remember going to events like Maker Faire and meeting lots of people who were excited and passionate about making. But it wasn’t a very diverse group and a lot of people were being professional makers by rediscovering traditional arts like learning how to ferment, learning how to sew. These are things in our community that people have as part of their traditions. Sewing in particular. Women of color get paid below minimum wage. And then you have people at Maker Faire who have learned sewing relatively recently who are making a decent living doing it but probably with less skill. I was really struck by the disparity there. Making felt like a rebranding of something that a lot of communities of color already do.''
\end{quote}

As Elena pointed out, members of these spaces often sought to reclaim hacking and making as a way of surviving; feminist and anti-racist spaces for creating in fact superseded the mainstream maker movement which in turn served to invisibilize the many contributions of women, people of color, and people with marginalized gender identities in innovation and entrepreneurship. With scarce resources, marginalized communities often had to innovate and improvise as a means of survival rather than a hobby, and many feminist makerspace founders recognized this.
 
Gene expressed a similar sentiment. Their makerspace was also POC-led, and they defined maker activities as coming from a place of necessity:

\begin{quote}
    ``When I think about makers, I think about my fellow Chicanos. My friend from high school who was too broke to build his own computer case so he built one out of wood.''
\end{quote}

In this context, making and hacking emerged from necessity and limited resources rather than from a desire to fit into a trendy maker culture. In this way, feminist makerspaces were seen as a way to create a shared space for marginalized community members to access a pool of resources, equipment, expertise, and mutual aid, not necessarily for the purpose of showcasing flashy tech projects, but instead as a means of ensuring the collective survival of marginalized makers.
 
\change{\textbf{\textit{Autoethnographic reflection. }}During the founding of our feminist makerspace in a cooperatively run building, POC tenants who were involved in racial justice organizing in the city proposed a decision-making structure that would give additional weight to votes made by POC in monthly tenant meetings. In these monthly meetings, tenants decided on issues such as rent rates, improving the accessibility of the building, parking, building maintenance, and event schedules. During one of these monthly meetings, the tenants agreed to adopt this weighted voting structure and moving forward, it became the policy for decision-making during all building-wide meetings.}


\section{Discussion}
\change{The cyclical nature of inclusion efforts in tech reveals the limits of corporate diversity initiatives, where gains are often followed by backlash. 
Over the past decade, feminist and antiracist interventions have ranged from advocacy within mainstream institutions to the creation of autonomous spaces that center marginalized communities. The rise of feminist hackerspaces and makerspaces in the mid-2010s exemplified a grassroots response to exclusionary cultures, offering alternative models for technical skill-building, governance, and collective care. However, as the industry retreats from diversity, equity, and inclusion (DEI) commitments, dismantling ethics teams, and reinstating exclusionary corporate values, these alternative spaces remain crucial. The recurring erosion of institutional DEI efforts underscores their precarity, reinforcing the need for community-driven approaches to equity and inclusion.}

\change{Rather than viewing feminist makerspaces and similar counterinstitutions as peripheral to the tech industry, we instead view these spaces as enacting alternative governance models that redistribute power, resources, and labor to sustain inclusive and accountable tech cultures. The industry’s present day retreat from DEI, along with its failure to address deep-seated inequality, reinforces the need for spaces that prioritize solidarity over corporate survival. As this paper argues, the endurance of feminist makerspaces and advocacy networks not only challenges the structural inertia of mainstream tech but also offers a roadmap for prefigurative politics, one that foregrounds mutual aid, collective empowerment, and long-term self-determination. In the remaining sections, we examine three key dimensions of sustaining these initiatives: stewardship, which considers the challenges of maintaining institutional knowledge and leadership within volunteer-driven spaces; solidarity, which explores alliances with local organizations and social justice coalitions; and scale, which interrogates the tension between deep community investment and the pressures to expand through institutional funding. Together, these themes illustrate how feminist makerspaces resist the structural inertia of mainstream tech while charting alternative pathways for more just technological futures.}

\subsection{The Necessity of Long-Term Stewardship}
\change{Feminist makerspaces—like many community-driven, justice-oriented tech projects—require a form of long-term stewardship that is rarely resourced, discussed, or supported in dominant narratives of innovation. Although the field of HCI and CSCW has robustly examined the creation of alternative sociotechnical infrastructures, less attention has been paid to what Le Dantec and Fox \cite{le2015strangers} have described as “work to keep the work going”. In the context of feminist makerspaces, this includes the ongoing labor of maintaining relationships, resources, knowledge infrastructures, and legitimacy amid oppressive cultures.}

\change{Several recurring challenges point to the need for sustained infrastructural care and stewardship. First, leadership instability and inexperience among board members can undermine organizational continuity. Unlike larger institutions, many feminist makerspaces rely on volunteers or part-time leaders who possess limited institutional knowledge about the space. When these individuals cycle out—a frequent occurrence due to competing life responsibilities and volunteer burnout—they often take with them critical expertise in areas such as grant-writing, technical maintenance, and everyday operational practices. As a result, these spaces become vulnerable to collapse not from a lack of interest, but from a lack of stewardship. In our study, several feminist makerspace founders departed shortly after launch, leaving legal and financial responsibilities to newly recruited board members unfamiliar with the space’s history or infrastructure. The absence of succession planning and governance continuity rendered these organizations particularly fragile, underscoring how the ideals of stewardship can falter without sustained infrastructural support.}

\change{Second, the absorption—or disappearance—of community-based initiatives into larger institutions can complicate questions of sustainability. 
Consider the example of OpenStreetMap (OSM), which began as a grassroots effort to build an open, editable map of the world. In order to sustain the project, major tech companies like Amazon and Meta became involved, contributing resources but also influencing priorities~\cite{chauhan2024value}. As a result, the project shifted, as corporate involvement reshaped community-led governance, and community-driven goals were subsumed into institutional agendas.
This echoes findings from our study, where the first author's decision to accept corporate funding in the early years of the makerspace-a decision made because of a lack of stewardship experience-enabled short-term growth but sparked internal conflict and estranged members committed to its anticapitalist mission. As with OSM, the support outside institutional investment challenged the community’s ability to preserve its founding values. Ultimately,  a commitment to collective stewardship is essential to handling these decisions with care — balancing tradeoffs towards sustaining alternative technology production.}

\change{Relatedly, efforts to sustain community-based participatory research (CBPR) are rarely designed with long-term stewardship in mind. As Kotturi, Hui et al. \cite{kotturi2024sustaining} argue, existing CSCW literature often focuses on the launch and short-term impacts of CBPR collaborations, but under-theorizes what happens when the formal research ends and the responsibility to maintain tools, practices, or relationships falls back on already overburdened communities. Feminist makerspaces often operate in these interstitial zones, where commitments to democratic design and care ethics clash with a lack of stable infrastructure.}

\change{Finally, creators of countertechnologies, those building viable alternatives to extractive platforms, face what Tandon et al. \cite{tandon2022hostile} describe as ``hostile ecologies'': restricted access to funding, inequitable regulatory frameworks, diminished design agency, and community partners with limited bandwidth or divergent priorities. These dynamics are familiar to those working on decentralized gig worker platforms, such as projects that Irani and Silberman~\cite{irani2013turkopticon}, and Calacci~\cite{calacci2022organizing} and others have explored. We observed similar forms of hostility in our study. As Gene’s account of displacement highlights, feminist makerspaces must be able to navigate hostile urban ecologies where rising rents and gentrification threaten their physical survival. Internal dynamics, too, can undermine collective care—as seen in the conflict with Dana, a Black trans woman whose grassroots outreach was dismissed by a founding member in favor of more “professional” approaches.  
Building and maintaining counterinstitutional systems requires not only technical expertise but persistent organizing, trust-building, and institutional repair, all of which are forms of stewardship. In this context, stewardship is not merely about keeping servers online or meeting grant deliverables. It is about cultivating the conditions under which feminist tech infrastructures can survive, through leadership succession planning, distributed knowledge sharing, collective ownership models, and alliances with other institutions of care. The future of feminist makerspaces depends not only on the moment of their creation but also on our collective willingness to sustain them.}

\subsection{Solidarity over Scale, in Order to Build Counterspaces}
\change{Throughout our interviews, founders and members of feminist makerspaces returned to a central question: How do we sustain our work in ways that center care, reciprocity, and collective survival? Rather than viewing sustainability as an internal organizational challenge or a question of longevity alone, participants reframed it through an interconnected lens of mutual aid and solidarity. Drawing on Dean Spade’s \cite{spade2020mutual} work, we understand mutual aid as the radical act of meeting people’s immediate needs through collective care and bottom-up organizing, rather than relying on institutional or philanthropic gatekeeping. As seen in efforts to keep tenant rents below market rate to prevent displacement, feminist makerspaces often rely on collective care and reciprocal support rather than external gatekeepers. Feminist makerspaces, then, are not only sites of technical production or education—they are critical infrastructures for redistributing resources and building interdependence across communities.}

\change{This orientation resonates with emerging conversations within the CSCW community around the role of community-based organizations in sustaining justice-oriented work. Research has shown that such organizations often act as vital nodes in larger networks of care, support, and mobilization \cite{haesler2021stronger, biorn2018building}. In these contexts, sustainability is not just about financial or organizational continuity, but also about cultivating long-term relationships and mutual obligations across diverse constituencies \cite{Mouffe_2022}. Participants in our study emphasized that responding to the evolving needs of their communities requires an ongoing commitment to redistribution—not just of funds, but of time, labor, space, and social support. This includes providing stipends for staff and volunteers, offering meals, childcare, and transportation, and continuously adapting programming to reflect the lived realities of those most impacted by intersecting forms of oppression. These efforts mirror broader solidarities found in movements for community land trusts \cite{tran2022careful} and the solidarity economy \cite{dillahunt2015promise, vlachokyriakos2017hci}, which reject capitalist logics of scarcity and competition in favor of collective wealth and shared governance.}

\change{By centering mutual aid and solidarity, feminist makerspaces have begun to articulate a vision of sustainability grounded not in scale, institutional prestige, or corporate partnerships, but in the everyday work of showing up for one another \cite{Incite_2017}. This approach demands that feminist makerspaces remain responsive to current social conditions, and that they engage in collaborative struggle with other social justice organizations as part of a broader fabric of resistance and care.}

\subsubsection{Feminist Makerspaces as Prefigurative Counterspaces}
\change{This paper has examined feminist makerspaces as counterspaces that challenge the dominant logics of mainstream tech institutions and the broader makerspace movement. These spaces do not simply offer alternative tools or techniques for making; rather, they work to reconfigure the cultural, social, and political conditions under which making happens. In doing so, they respond to calls for prefigurative design~\cite{asad2019prefigurative}—a practice rooted in the belief that justice-oriented futures must be enacted in the present through community-driven design processes.}

\change{Feminist makerspaces, particularly those that have centered the needs and leadership of marginalized community members, embody this prefigurative ethos by refusing the extractive, individualistic, and profit-driven values that often characterize mainstream tech. Instead, they foster collective forms of knowledge production, care, and solidarity. Following Asad’s framework~\cite{asad2019prefigurative}, these spaces can be understood as sites where community members name harms, prioritize their own needs, and build the infrastructural and emotional support necessary for healing and autonomy. In other words, these spaces do not just resist dominant institutions—they build new ones. As we saw in our findings, this included leadership and decision making by those who are the most marginalized, inviting neighbors from nearby shelters into programming, and creating onboarding structures that helped new members feel empowered to participate, even without prior experience.}

\change{Moreover, the sustained and regionally networked feminist makerspaces we observed contribute to what scholars have called the solidarity economy~\cite{spade2020mutual,vlachokyriakos2017hci}. These economies operate outside of or adjacent to the market, creating systems of mutual aid, shared labor, and distributed governance that prioritize care and survival over competition and efficiency. This contribution is particularly critical given how mainstream tech continues to reproduce structural inequalities—excluding women and people of color from hiring pipelines, promotion ladders, and even from the historical narratives of technological innovation \cite{rosner2018making, nakamura2014indigenous, hicks2017programmed}. Feminist makerspaces reclaim and revalue these lineages of making, affirming them not only as legitimate but as foundational \cite{rosner2016legacies}.}

\change{Importantly, the work of feminist makerspaces is not merely oppositional. While they resist and critique dominant systems, they also offer affirmative visions of what technology work can be. Making and hacking, in this context, become practices of relation, repair, and refusal—practices that are deeply grounded in shared values and collective imagination. As counterinstitutions, these spaces illuminate the possibilities for building just and liberatory futures, not in abstraction but through the everyday labor of community-making. In recognizing feminist makerspaces as prefigurative counterspaces, CSCW scholars are invited to take seriously the infrastructural, affective, and political labor involved in building such alternatives. These spaces challenge us to reimagine not only what gets made, but how, by whom, and toward what ends.}


\section{CONCLUSION}
Throughout this paper, we have examined the everyday social practices through which feminist makerspaces sustain themselves, promote alternative forms of technology production, and contribute to regional social justice efforts. By highlighting the coalition-building strategies adopted in some spaces, we point to a path forward for alternative tech cultures—one grounded in feminist commitments to collective survival and solidarity. In particular, we show how a solidarity-based approach to advancing gender and racial equity in technology can help makerspace founders deepen ties with aligned organizations. \change{While philanthropic and corporate funders often emphasize scale—urging expansion of programs and outreach—founders stressed that sustainability lies instead in cultivating fewer, deeper relationships with members and community partners. Rather than prioritizing growth, these spaces find resilience in embedding their work within broader movements for justice. Feminist makerspaces can not transform the tech sector alone, but through sustained, solidaristic relationships rooted in feminist values, they offer a powerful model for reimagining technology production as collective, justice-oriented, and community-embedded work.}

\begin{acks}
To all of the feminist makerspace founders and members who shared their time and expertise with us. Thank you for doing this work and showing that another world is possible.
\end{acks}

\bibliographystyle{ACM-Reference-Format}
\bibliography{main}

\end{document}